\documentclass[twocolumn,amsmath,showkeys,amssymb,superscriptaddress,longbibliography,floatfix, pre]{revtex4-2}

\pdfoutput=1

\usepackage{bm}
\usepackage{yfonts}
\usepackage{graphicx,epsfig}
\usepackage{amsmath,amssymb,mathtools}
\usepackage[rgb]{xcolor}
\usepackage{hyperref}
\usepackage{braket}
\def\be{\begin{equation}}
\def\ee{\end{equation}}

\def\bc{\begin{center}}
\def\ec{\end{center}}
\def\bea{\begin{eqnarray}}
\def\eea{\end{eqnarray}}

\begin{document}

\title{Thermodynamics of the  Gravity from Entropy theory}

\author{Ginestra Bianconi}
\email{ginestra.bianconi@gmail.com}
\affiliation{School of Mathematical Sciences, Queen Mary University of London, London, E1 4NS, United Kingdom}

\begin{abstract}
The Gravity from Entropy (GfE) action posits that gravity that is fundamentally given by the information encoded in the interplay between matter and geometry. The GfE Lagrangian is given by the Geometric Quantum Relative Entropy (GQRE) between the physical metric and the metric induced by matter and curvature, leading to modified gravitational field equations with an emergent dynamical  effective dark energy term, which reduce to Einstein’s equations in the low energy, small curvature limit. Adopting a thermodynamic viewpoint, we identify the GfE energy density with this emergent effective dark energy term. For homogeneous and isotropic FRW spacetimes, we show that  GfE universes admit a thermal description: locally, they are characterized by $k$-temperatures and $k$-pressures satisfying a first law of GfE thermodynamics. In the low energy, small curvature regime with perfect-fluid matter and radiation, GfE solutions are well approximated by Friedmann cosmologies. We show that, while the total GQRE per unit volume does not increase, consistent with its nature as a relative entropy, the total entropy of GfE universes is non-decreasing in time. These results provide a thermodynamic interpretation of GfE cosmologies and of general relativity (GR) itself,  recovered in the low energy, small curvature limit of the theory, offering a framework to reconcile local order and complexity with the global increase of entropy in the universe.
\end{abstract}
\maketitle

 \section{Introduction}
 
It is generally believed that the early universe corresponds to a state of small entropy and that the universe evolves toward a high entropy configuration. However we still miss a consensus on the  theory that is able to quantify the geometric degrees of freedom reconciling this picture with gravity on a solid statistical mechanics basis.

The recently proposed Gravity from Entropy (GfE) theory \cite{bianconi2025gravity,bianconi2025quantum,bianconi2025beyond} posits that gravity is fundamentally a  statistical mechanics theory capturing the information  encoded in the interplay between the  geometric degrees of freedom of spacetime and the matter fields. The present work develops the thermodynamics of the GfE theory for homogeneous and  isotropic spacetimes and sheds new light on the relation between cosmology and thermodynamics.
Specifically here it is  shown that while the total entropy of the universe increases with time in the GfE theory, the entropy per unit volume decreases. {These results frame both GfE cosmologies and general relativity recovered  in the low energy, small curvature limit of the theory, within a thermodynamic perspective, reconciling local order and complexity with the global growth of entropy in the universe.}

Since the discovery of the thermodynamics of black holes \cite{bekenstein1973black,bekenstein1974generalized,hawking1975particle,witten2025introduction,kubizvnak2017black,yokokura2025black} 
 the deep connection between gravity and statistical mechanics has attracted large scientific attention. 
 Further insight into these profound relations has been provided by early results on the entropy and the temperature of de Sitter space \cite{gibbons1977cosmological,bunch1978quantum,birrell1984quantum} that have triggered  significant attention because of their profound implications \cite{bousso2002conformal,bousso2002adventures}. Interestingly, the horizon entropy, including Wald entropy and its extensions has been interpreted  in terms of the Noether charge of modified gravity actions \cite{wald1993black,iyer1994some,jacobson1994black,sasa2016thermodynamic}. This entropy allows to capture corrections to the area law derived from higher curvature terms in the Lagrangian.
All together, these findings demonstrate that the relation between gravity and statistical mechanics  involves  spacetimes with a horizon other than black holes. Therefore for several decades, a longstanding challenge in theoretical physics has been  to establish the profound connections between statistical mechanics,  information theory, and gravity.
 So far, different entropic gravity approaches have been proposed \cite{jacobson1995thermodynamics,cai2005first,carroll2016entropy,jacobson2016entanglement,verlinde2011origin,callan1994geometric,chirco2014spacetime,chirco2010nonequilibrium,padmanabhan2010thermodynamical,dorau2025quantum}
  in the literature, which take as the fundamental starting point the area law and more in general horizon entropy.
The horizon entropy also plays an important role in the holographic theory \cite{hooft2001holographic,susskind1995world,swingle2012entanglement},  relating information theory, entanglement entropy \cite{ryu2006aspects,blanco2013relative,casini2011towards,nishioka2009holographic,faulkner2013quantum}  and gravity. From the statistical mechanics point of view, however,  the quest for a theory that captures the elementary degrees of freedom of geometry is ongoing. From  the cosmological point of view, the search for modified area laws beyond what is already captured by the Wald entropy, has led to the discussion of  generalized entropies in cosmology \cite{barrow2020area,kaniadakis2002statistical,tsallis2013black,tsallis2025extensive,heymans2025entropy,quevedo2007geometrothermodynamics}. 
Their potential ability to interpret cosmological observations is currently a very active subject of research
~\cite{capozziello2018information,capozziello2022thermodynamic,capozziello2023black,luciano2024kaniadakis,capozziello2025barrow,nojiri2022nonextensive,nojiri2022early}.
 {More broadly the general problem of reconciling our understanding on gravity with the second principle of thermodynamics is a classic topic in General Relativity (GR) and cosmology \cite{padmanabhan2010thermodynamical,baumann2022cosmology}. However, in standard GR, the entropy is primarily attributed either to matter and  radiation, or to horizons, and geometric degrees of freedom ~\cite{clifton2013gravitational} but not directly on the interplay between geometric degree of freedom and matter fields like in the GfE theory.
}

In addition to the discussion on the relation between gravity and statistical mechanics, another very active research line concerns the lively discussion on the ultimate nature of gravity. Traditionally this is discussed within the different approaches to quantum gravity ranging from strings~\cite{polchinski1998string} and loop quantum gravity~\cite{Rovelli} to asymptotic safety~\cite{Rahmede,eichhorn2019asymptotically} and causal dynamical triangulations~\cite{CDT}.
 Recently the field has involved also the proposal of experiments that build on the theory of entanglement to assess whether the nature of gravity is quantum 
\cite{bose2017spin,marletto2017gravitationally,bose2025massive} and the vibrant scientific debate that these works have generated.
Thus this growing research line points out the key role that entanglement will have in a full quantum gravity theory.

The recently proposed GfE theory \cite{bianconi2025gravity,bianconi2025quantum,bianconi2025beyond} stands out among the entropic gravity approaches as it proposes an alternative action for gravity, the GfE action, that reduces to the Einstein-Hilbert action in the low energy, small curvature limit. This action is associated with a Lagrangian given by the Geometrical Quantum Relative Entropy (GQRE), which is defined under the assumption that the metrics associated with spacetime can  actually be treated as quantum operators. The notion of GQRE on which the GfE theory is based  provides a new interpretation of gravity as a relational theory between the true metric and the metric induced by the matter fields and curvature. Moreover the GQRE differs conceptually from horizon entropies as it is local and volumeric and  is  not derived from horizons. Interestingly, the GQRE  reveals significant connections with the Araki entropy \cite{araki1975relative,araki1999mathematical,ohya2004quantum} used in the theory of quantum local operators~\cite{sorce2023notes,jensen2023generalized} 
  and proposed by Witten as the promising approach  for capturing quantum entanglement avoiding ultraviolet divergences \cite{witten2018aps}.  
 The GfE action, so far developed in first quantization, leads to the GfE modified gravity equations \cite{bianconi2025gravity} which depend on an emergent field, called the G-field which dresses the metric and depends both on the matter fields  and on the geometry.
 Specifically, these equations reveal the emergence of a dynamical  {effective dark energy term}  $\Lambda_{\mathcal{G}}$  which  depends exclusively on the G-field.
Therefore, the GfE approach opens new scenarios  for exploring cosmological implications for dynamics of dark energy~\cite{copeland2006dynamics,bahamonde2018dynamical}, the early universe  inflationary behaviour \cite{thattarampilly2025inflation} and    {leads to testable predictions for gravity beyond general relativity  that have started to be explored in  Ref.\cite{thattarampilly2026spherically}. Thus GfE theory  might be tested  against present and future cosmological and astrophysical observations} \cite{berti2015testing,barausse2020prospects,di2025cosmoverse}.

 {Given that the GfE posits gravity as the statistical mechanics theory encoding the information present in the interplay between the geometric degree of freedom and matter fields, } an important open question is the characterization of its associated thermodynamics.

Progress in this direction has been already made in Ref.\cite{bianconi2025quantum} where it was demonstrated that the GfE action   associated with the Schwarzschild black hole which indicates the GfE entropy, obeys the area law.  This is a key result that reveals how   the area law for Schwarzschild black holes can be derived from the first principles of the GfE theory without relying on general relativity or on any holographic assumption.

Another  fundamental statistical mechanics question is to establish whether the GfE action is consistent with uniform and homogeneous space. This question has been tackled  in Ref. ~\cite{bianconi2025beyond}    within a simplified setting called the warm-up scenario of the GfE theory. Indeed, in Ref.~\cite{bianconi2025beyond} it is pointed  out that in this simplified setting homogeneity and uniformity can be realized in large part of spacetime but still the maximization of the GfE entropy can be consistent with the local preservation of non trivial structures.  On the basis of the mentioned results our belief is that also the full GfE theory might not simply lead to maximal uniformity and homogeneity, differently from the results obtained in more traditional settings  \cite{turok2024gravitational,boyle2024thermodynamic}.
 
 Leaving further investigation of these aspects of the GfE theory to further work, here we focus on the thermodynamics of the GfE theory. Specifically, in this work we derive  the GfE thermodynamics which will allow us to  establish whether and by which mechanism the entropy of the universe increases in time, and whether it is possible to associate a temperature and  a pressure to the geometric degrees of freedom of spacetime.

 To this end, we identify the internal energy density with the emergent  {effective dark energy term} and the local entropy density with the GQRE and we derive the GfE thermodynamics.
We show that locally GfE spacetimes are thermal and they are associated with the $k$-temperatures and the $k$-pressures that depend on the nature and the order of the geometric degrees of freedom that are considered. Since GfE is a higher-order theory of gravity \cite{kuipers2025quantum} we have a distinct temperature and a pressure for the  scalar degrees of freedom, and the timelike and spacelike vector and the bivector degrees of freedom.
These thermodynamic quantities are related by the first law of thermodynamics of the GfE theory.
In order to provide a concrete example of the implications of this theory, we approximate GfE universes  in the low energy, small curvature limit with Friedmann universes  and we provide a criterion to assess under which conditions this approximation is valid for the GfE theory.
As long as this criterion is satisfied we show that the total GQRE  and the local energy of the universes per unit volume go to zero but the volume contribution diverges with time.
This combined effect leads to the divergence of  the total entropy with time  while the total energy saturates to a constant for radiation and matter dominated universes. 
We consider also the implications of this theory for the de Sitter space and show that for a small Hubble parameter $H$ the total entropy associated with the de Sitter space causal diamond scales as $H^{-2}$.

\section{ The Gravity from Entropy theory}
\subsection{The GfE action}
 We consider a four-dimensional ($d=4$) spacetime $\mathcal{K}$ with signature $(-1,1,1,1)$ and metric $g_{\mu\nu}$ associated with a Levi-Civita connection. Moreover units $c=\hbar=1$ are adopted throughout the paper, while the Planck length is indicated as $\ell_P$.
 The Gravity from Entropy theory (GfE)  \cite{bianconi2025gravity} treats the metrics  associated with spacetime as quantum operators and proposes that the Lagrangian for gravity is given by their associated Geometric Quantum Relative Entropy (GQRE). Thus gravity is the outcome of the interplay between the actual true metric of the manifold and the metric induced by the matter fields and the curvature. 
 The resulting GfE theory stems from the GfE action $\mathcal{S}$ 
 \cite{bianconi2025gravity}  given by 
 \bea
 \mathcal{S}=\frac{1}{\ell_P^4}\int \sqrt{-|g|}\mathcal{L}d^4{\bf r}
 \label{ActionGfE}
 \eea
where the Lagrangian $\mathcal{L}$, is given by the GQRE between the true  metric $\tilde{g}$ associated with spacetime  and the metric $\tilde{\bf G}$ induced by the matter fields and curvature. Specifically,  the GfE Lagrangian is given by
 \bea
  \mathcal{L}=-\mbox{Tr}_F\ln({\bf \tilde{G}}\tilde{g}^{-1}), 
  \label{LGfE}
 \eea
 where here  the trace is the trace of the flattened tensor and ${\bf \tilde{G}}\tilde{g}^{-1}$ is assumed to be defined positive. 
 
 The true higher-order metric $\tilde{g}$ associated with spacetime  is given by 
 \bea
\tilde{g}=1\oplus g_{\mu\nu}dx^{\mu}\otimes dx^{\nu}\oplus [g_{(2)}]_{\mu\nu\rho\sigma}dx^{\mu}\wedge dx^{\nu}\otimes dx^{\rho}\wedge dx^{\sigma},\nonumber
\eea
with
\bea
[g_{(2)}]_{\mu\nu\rho\sigma}=\frac{1}{2}(g_{\mu\rho}g_{\nu\sigma}-g_{\mu\sigma}g_{\nu\rho}).
\eea
  The higher-order metric induced by the matter field and curvature $\tilde{\bf G}$ has a  structure similar to the true higher-order metric, i.e.
\bea
\tilde{\bf G}&=&G_{(0)}\oplus [G_{(1)}]_{\mu\nu}dx^{\mu}\otimes dx^{\nu}\nonumber \\ &&\oplus [G_{(2)}]_{\mu\nu\rho\sigma}dx^{\mu}\wedge dx^{\nu}\otimes dx^{\rho}\wedge dx^{\sigma}.
\eea
In the GfE theory \cite{bianconi2025gravity} this metric $\tilde{\bf G}$ is given by 
\bea
\tilde{\bf G}=\tilde{g}+\alpha\tilde{\bf M}-\beta\tilde{\bm{\mathcal{R}}},
\label{Gtilde}
\eea
where $\alpha=\alpha^{\prime}\ell_P^{4}$, $\beta=\beta^{\prime}\ell_P^2$, with $\alpha^{\prime}, \beta^{\prime}$ indicating dimensionless positive constants. As we discuss in the next sections these constants can be constrained by imposing that the GfE action reduces to the Einstein-Hilbert action plus matter action in the low-energy,  small-curvature limit.
In Eq.(\ref{Gtilde}), $\tilde{\bf M}$ indicates the contribution of the matter fields and is given by 
\bea
\tilde{\bf M}&=&M_{(0)}\oplus [M_{(1)}]_{\mu\nu}dx^{\mu}\otimes dx^{\nu}\nonumber \\ &&\oplus [M_{(2)}]_{\mu\nu\rho\sigma}dx^{\mu}\wedge dx^{\nu}\otimes dx^{\rho}\wedge dx^{\sigma},
\eea
while $\tilde{\bm{\mathcal{R}}}$ is given by the direct sum of the Ricci scalar, Ricci tensor, and Riemann tensor, i.e. 
\bea
\tilde{\bm{\mathcal{R}}}&=&a_0R\oplus a_1R_{\mu\nu}dx^{\mu}\otimes dx^{\nu}\nonumber \\&&\oplus  R_{\mu\nu\rho\sigma}dx^{\mu}\wedge dx^{\nu}\otimes dx^{\rho}\wedge dx^{\sigma},\label{Rcurly}
\eea
 {with $a_0\geq 0,a_1\geq 0$  adimensional parameters. Here and in the following we take $a_0=a_1=1$.}
The term  $\tilde{\bf M}$ is designed to include potentially all matter fields. For explicit expressions in the case of  bosonic matter fields and Abelian gauge field we refer the reader to Ref.\cite{bianconi2025gravity}.

Interestingly, we observe that while the GfE Lagrangian $\mathcal{L}$ given by Eq.(\ref{LGfE}) is given by the GQRE, this quantity can be also interpreted as a Gibbs-Maxwell entropy as we have
\bea
\mathcal{L}=\ln W
\eea
where $W$ quantifies the number of degrees of freedom encoded in the metric  $\tilde{\bf G}$ that can be codified by $\tilde{g}$ and is given by
\bea
W=G_{(0)}^{-1}(\mbox{det}(g_{(1)}{\bf G}_{(1)}^{-1}))(\mbox{det}(g_{(2)}{\bf G}_{(2)}^{-1})).
\eea
Thus the action $\mathcal{S}$ given by Eq.(\ref{ActionGfE}) can be interpreted as the entropy quantifying the geometric degrees of freedom of spacetime.
As we will demonstrate in this work, the GQRE Lagrangian $\mathcal{L}$ preserves some aspects of the standard quantum relative entropy \cite{vedral2002role} which measure distinguishability between quantum states and is not increasing while the action $\mathcal{S}$ of GfE is actually interpretable as an entropy as is not decreasing for GfE universes in the low-energy,  small-curvature limit.
We note that since the GfE Lagrangian is a scalar quantity the action in (\ref{ActionGfE}) is diffeomorphism invariant. However the GfE action goes beyond the minimal coupling and the matter and the geometric degree of freedom cannot be treated in isolation. Moreover,  the GfE Lagrangian treats naturally also metrics associated with horizons  {while further investigations will need to establish whether true singularities such as the Big Bang and the black hole singularities are dynamically avoided in the GfE theory. }
 {The very precise structure of GfE action is dictated by statistical mechanics principles, and while it might look as a strong constrain on the nature of nonlinear terms of this modified gravity theory,  its functional form embodies physical meaning in terms of the GQRE.}
\subsection{The G-field}
The  GfE Lagrangian given by Eq.(\ref{LGfE}) is non-linear in the curvature, however this non-linear dependence involves  exclusively  the quantity 
\bea
\tilde{\bm \Theta}=\tilde{\bf G}\tilde{g}^{-1}.
\eea
 Thus the GfE Lagrangian can be expressed by enforcing these constraints with Lagrangian multipliers $\tilde{\bm{\mathcal{G}}}$ leading to the following expression for the Lagrangian \cite{bianconi2025gravity}
\bea
{\mathcal{L}}=-\mbox{Tr}_F\ln{\tilde{\bm \Theta}}-\mbox{Tr}_F\left[\tilde{\bm{\mathcal{ G}}}(\tilde{\bf G}\tilde{g}^{-1}-\tilde{\bm\Theta })\right].
\label{L2GfE}
\eea
In the GfE theory  the G-field $\tilde{\bm{\mathcal{G}}}$ is treated as  a physical and possibly measurable field, which can be related to $\tilde{\bm\Theta}$ by
\bea
\tilde{\bm{\mathcal{G}}}=\tilde{\bm\Theta}^{-1}.
\label{Gfield}
\eea
Thus the GfE theory assumes that the G-field can be treated like other notable Lagrangian multipliers  in statistical mechanics which are physical and measurable such as the temperature and the chemical potential. 
 {Substituting Eq.(\ref{Gfield}) into Eq.(\ref{L2GfE})  we obtain that the Lagrangian $\mathcal{L}$ can be written as {\bea
{\mathcal{L}}=\mbox{Tr}_F\ln{\tilde{\bm{\mathcal{G}}}}-\mbox{Tr}_F\left[\tilde{\bm{\mathcal{ G}}}\tilde{\bf G}\tilde{g}^{-1}-\tilde{\bf I}\right],
\label{LGfExx}
\eea}
where $\tilde{\bf I}$ indicates the topological identity.
Therefore using the explicit expression for $\tilde{\bf G}$ given by  Eq.(\ref{Gtilde}) the Lagrangian} decouples into a gravitational Lagrangian $\mathcal{L}_G$ and a matter Lagrangian  $\mathcal{L}_M$, i.e.
\bea
{\mathcal{L}}=\beta\mathcal{L}_G+\alpha\mathcal{L}_M
\label{LGfEx}
\eea
The gravitational  Lagrangian and the matter Lagrangian involve contraction with a {\em dressed metric} $\tilde{g}_{\mathcal{G}}$ given by 
\bea
\tilde{g}_{\mathcal{G}}={\tilde{\bm{\mathcal{G}}}}^{-1}\tilde{g}
\eea
and includes an  {{\em emergent effective dark energy term}} $\Lambda_{\mathcal{G}}$ that depends exclusively on the G-field 
\bea
\Lambda_{\mathcal{G}}=\frac{1}{2\beta}\mbox{Tr}_F\left[\tilde{\bm{\mathcal{G}}}-\tilde{\bf I}-\ln \tilde{\bm{\mathcal{G}}}\right],
\label{LambdaG}
\eea
where $\tilde{\bf I}$ indicates the higher-order identity (see for details \cite{bianconi2025gravity}).
Specifically the gravitational  Lagrangian $\mathcal{L}_G$ is  given by
\bea
\mathcal{L}_G&=& \left({\mathcal{R}}_{\mathcal{G}}-2\Lambda_{\mathcal{G}}\right)\eea
where the dressed Ricci scalar ${\mathcal{R}}_{\mathcal{G}}$ is given by 
\bea
{\mathcal{R}}_{\mathcal{G}}=\mbox{Tr}_F \tilde{\bm{\mathcal{R}}}\tilde{g}_{\mathcal{G}}^{-1},
\eea
 while the matter Lagrangian $\mathcal{L}_M$ is given by
\bea
\mathcal{L}_M&=&- \mbox{Tr}_F \tilde{{\bf M}}\tilde{g}_{\mathcal{G}}^{-1}.
\eea
 {Note that the  {\em emergent  {effective dark energy term}} arises from the nonlinear structure of the GfE action, a mathematical feature that is suggesting its   {interpretation as a dynamical cosmological constant}. As we will see not only in the action but also in the modified Einstein equations derived from the GfE action, this contribution appears  {to play the role of effective dark energy which is reminiscent of }  a cosmological constant,  {with the important difference that it is actually dynamically generated by the G-field rather than imposed as a fundamental parameter of the theory.}
From this discussion it follows that  giving a physical interpretation to the G-field provides a reformulation of the GfE action  that is linear in the curvature and the G-field and allows us to cast the GfE action into the sum of a gravitational Lagrangian and a matter Lagrangian. Note however that the presence of the G-field in both Lagrangians  implies that matter and geometry cannot be treated independently.
In this work we will argue that the $G$-field must be interpreted as a dynamical field accounting for the internal energy density of the GfE universes (to be identified in the emergent  {effective dark energy term} $\Lambda_{\mathcal{G}}$). 
Therefore the G-field  is key to the evolution of the thermodynamic quantities of the GfE theory.
\subsection{The GfE modified gravity equations}
The GfE action $\mathcal{S}$ gives rise to the GfE modified gravity equations  for the metric $\tilde{g}$  given by 
\bea
{R}^{\mathcal{G}}_{(\mu\nu)}-\frac{1}{2}{g}_{\mu\nu}\Big(\mathcal{R}_{\mathcal{G}}-2\Lambda_{\mathcal{G}}\Big)+{{\mathcal{D}}}_{(\mu\nu)}=\frac{\alpha}{2\beta}\mathcal{T}_{\mu\nu},
\label{GfEeq}
\eea
where  $(\mu\nu)$ indicates the symmetrization of the indices, ${R}^{\mathcal{G}}_{\mu\nu}$ are the elements or the {\em dressed Ricci tensor} given by
\bea
{R}^{\mathcal{G}}_{\mu\nu}&=&{\mathcal{G}_{(0)}}R_{\mu\nu}+{[{\mathcal{G}_{(1)}}]}_{\mu}^{\ \rho}R_{\rho\nu}-{[\mathcal{G}_{(2)}]}_{\rho_1\rho_2\mu\eta}R_{\nu}^{\ \eta\rho_1\rho_2}\nonumber \\
&&+2{[{\mathcal{G}_{(2)}}]}_{\mu}^{\ \eta\rho_1\rho_2}R_{\rho_1\rho_2\nu\eta},
\eea
while ${{\mathcal{D}}}_{\mu\nu}$ are the elements depending on second derivatives of the G-field $\tilde{\bm{\mathcal{G}}}$ given by 
\bea
{{\mathcal{D}}}_{\mu\nu}&=&(\nabla^{\rho}\nabla_{\rho}g_{\mu\nu}-\nabla_{\mu}\nabla_{\nu}){\mathcal{G}_{(0)}}-\nabla^{\rho}\nabla_{\nu}{[\mathcal{G}_{(1)}]}_{(\rho\mu)}
\nonumber \\
&&+\frac{1}{2}
\nabla^{\rho}\nabla_{\rho}{[\mathcal{G}_{(1)}]}_{\mu\nu}+\frac{1}{2}\nabla^{\rho}\nabla^{\eta}{[\mathcal{G}_{(1)}]}_{\rho\eta}g_{\mu\nu}\nonumber \\
&&+\nabla^{\eta}\nabla^{\rho}{[\mathcal{G}_{(2)}]}_{\mu\rho\nu\eta}+\nabla^{\rho}\nabla^{\eta}{[\mathcal{G}_{(2)}]}_{\eta\mu\rho\nu}\nonumber \\
&&+\frac{1}{2}[\nabla^{\rho},\nabla^{\eta}]{[\mathcal{G}_{(2)}]}_{\rho\eta\mu\nu}.
\eea
Defining the matter action $\mathcal{S}_M$ as 
\bea
\mathcal{S}_M=\int\sqrt{-|g|}\mathcal{L}_M d^4{\bf r},
\eea
the dressed stress-energy tensor $\mathcal{T}_{\mu\nu}$ appearing in the GfE equations of motion (\ref{GfEeq}) is given by
\bea
\mathcal{T}_{\mu\nu}=\frac{-2}{\sqrt{-|g|}}\frac{\delta \mathcal{S}_M}{\delta g^{\mu\nu}}.
\eea
The investigation of the exact solutions of the GfE equations of motion has the potential to shed new light   in cosmology and has been argued in Ref. \cite{thattarampilly2025inflation} to lead to inflationary behavior without the need to introduce a scalar field.
Leaving the study of the exact solutions of the GfE equations to future works, here we focus on the thermodynamic properties associated with the GfE universes  focusing on the limit of low energy and small curvature.

\subsection{The General Relativity limit}
General Relativity is recovered in the first order approximation of the GfE theory  in the limit of low energy and small curvature, i.e. for $\alpha^{\prime}\ll 1,\beta^{\prime}\ll 1$.[Note that the linearization here is different from the conventional linearization in terms of a metric perturbation against e.g. a Minkowski background.]} In this limit the GfE Lagrangian reduces to a linear combination of the Einstein-Hilbert general relativity Lagrangian $\hat{\mathcal{L}}_{EH}$ and the matter field Lagrangian $ \hat{\mathcal{L}}_M$. Indeed   we obtain
\bea
\frac{1}{\ell_P^4}\mathcal{L}\simeq 3\beta^{\prime} \hat{\mathcal{L}}_{EH}+\alpha^{\prime} \hat{\mathcal{L}}_M,
\label{GR}
\eea
where 
\bea
\hat{\mathcal{L}}_{EH}&=&\frac{1}{\ell_P^2}R,\quad
\hat{\mathcal{L}}_M=-\mbox{Tr}_F\tilde{\bf M}\tilde{g}^{-1}.
\eea
 {This implies that  this linear approximation of the GfE Lagrangian $\mathcal{L}$ given by Eq.(\ref{LGfEx}) by putting $\Lambda_{\mathcal{G}}$ equal to zero, and in the remaining terms  retaining only the  zero-order approximation for the G-field. Therefore in this limit the $G$-field that dressed the metric reduces to  the higher-order identity, $\tilde{\bm{\mathcal{G}}}=\tilde{\bf I}$.}
The constants $\alpha^{\prime}$ and $\beta^{\prime}$ defined in Eq.(\ref{Gtilde}) for the full GfE theory, are therefore determined by requiring that Eq.(\ref{GR}) obtained in the low-energy,  small-curvature limit  lead to the Einstein equations. Thus this requirement implies that  the two dimensionless constants $\alpha^{\prime}$ and $\beta^{\prime}$ entering  the GfE action and defined in Eq.(\ref{Gtilde}) are not arbitrary but their values are constrained by the requirement that
\bea
\frac{\alpha^{\prime}}{3\beta^{\prime}}=16\pi. 
\label{ab}
\eea  
Therefore this result reduces the  {parameters} of the GfE theory to a single one. Therefore, by further specifying the theory, this strengthens its possible predictive power.

\section{Homogeneous and isotropic universes}
\subsection{The GfE theory for isotropic universes}
 Here and in the following we will consider homogeneous and isotropic models of the universe having the
Friedmann-Robertson-Walker (FRW) metric
\bea
ds^2=g_{\mu\nu}dx^{\mu}dx^{\nu}=-dt^2+a^{2}(t)\left[\frac{dr^2}{1-\kappa r^2}+r^2d\Omega^2\right]\nonumber
\eea
where $d\Omega^2=d\theta^2+\sin^2\theta d\phi^2$.
In this scenario the major quantity of interest in GfE is given by 
\bea
\tilde{\bf G}\tilde{g}^{-1}=\tilde{\bf I}-\tilde{\bf U}
\eea
where 
\bea
\tilde {\bf U}&=&U_{(0)}\oplus {[U_{(1)}]}_{\mu}^{\ \nu}dx^{\mu}\otimes dx_{\nu}\nonumber \\
&&\oplus {[U_{(2)}]}_{\mu\nu}^{\ \ \rho\sigma}dx^{\mu}\wedge dx^{\nu}\otimes dx_{\rho}\wedge dx_{\sigma}
\eea
Because of the homogeneity and isotropy of the FRW space the only non-zero flatted distinct eigenvalues of $\tilde{\bf U}$ defined in \cite{bianconi2025gravity} are 
\bea
u_0&=&U_{(0)}=\beta R-\alpha M_{(0)}\nonumber \\
u_1&=&{[U_{(1)}]}_0^{\ 0}=\beta R_0^{\ 0}-\alpha \left[M_{(1)}\right]_{0}^{\ 0}\nonumber \\
u_2&=&{[U_{(1)}]}_{i}^{\ i}=\beta R_i^{\ ji}-\alpha \left[M_{(1)}\right]_{i}^{\ i}\nonumber \\
u_3&=&2{[U_{(2)}]}_{0i}^{\ \ 0i}=2\left[\beta R_{0i}^{\ \ 0i}-\alpha \left[M_{(2)}\right]_{0i}^{\ \ 0i}\right]\nonumber \\
u_4&=&2{[U_{(2)}]}_{ij}^{\ \ ij}=2\left[\beta R_{ij}^{\ \ ij}-\alpha \left[M_{(2)}\right]_{ij}^{\ \ ij}\right],\label{tauk}
\eea
 where the curvature terms are given by 
\bea
R&=&6\left[\frac{\ddot{a}}{a}+\left(\frac{\dot{a}}{a}\right)^2+\frac{\kappa}{a^2}\right],\nonumber \\
R_0^{\ 0}&=&3\frac{\dot{a}}{a},\nonumber \\
 R_i^{\ i}&=&\left[\frac{\ddot{a}}{a}+2\left(\frac{\dot{a}}{a}\right)^2+2\frac{\kappa}{a^2}\right],\nonumber \\
R_{0i}^{\ \ 0i}&=&\frac{\ddot{a}}{a},\nonumber \\
 R_{ij}^{\ \ ij}&=&\left[\left(\frac{\dot{a}}{a}\right)^2+\frac{\kappa}{a^2}\right].
 \label{Rk}
\eea
Note that in the above Eqs.(\ref{tauk}) and (\ref{Rk}) repeated indices are not summed over.
 We indicate with $z_k$ the degeneracy of the eigenvalues $u_k$, defined in Eqs.(\ref{tauk}) thus we have $z_k=1$ for $k\in \{0,1\}$ while otherwise $z_k=3$.
The GfE Lagrangian   given by Eq.(\ref{LGfE}) can be written for homogeneous and isotropic universes as 
\bea
\mathcal{L}=-\sum_{k=0}^4 z_k\ln(1-u_k).
\eea
\subsection{The internal energy density $\mathcal{E}$ }
Defining an Hamiltonian formalism for gravity is notoriously challenging and connected to the problem of time~\cite{isham1993canonical,rovelli2004quantum,rovelli2015covariant}.
Here we take a full statistical mechanics approach  and we take as internal energy density $\mathcal{E}$ of the GfE theory the emergent  {effective dark energy term} $\Lambda_{\mathcal{G}}$ multiplied by $2\beta$, i.e.
\bea
\mathcal{E}=2\beta \Lambda_{\mathcal{G}}.
\eea
 Interestingly, this definition of the   internal energy density $\mathcal{E}$ of the GfE theory coincides with  the Legendre transformation of the GfE Lagrangian. Indeed since the GfE Lagrangian $\mathcal{L}$ is a  convex function of $u_k$, we can treat each contribution coming from a different flattened eigenvector independently and  consider the conjugate variable ${\mathcal{G}}_k$ of $u_k$ defined as
 \bea
{\mathcal{G}}_k=\frac{\partial \mathcal{L}}{\partial u_k}=\frac{1}{1-u_k}.
 \eea
 This expression thus reveals that ${\mathcal{G}}_k$ is the $k$-th component of the G-field. From this equation it follows that $u_k$ can also be expressed as 
 \bea
 u_k=1-{\mathcal{G}}_k^{-1}.
 \eea
 Here we show that the internal energy density $\mathcal{E}=\mathcal{E}(\boldsymbol{\mathcal{G}})$ is the Legendre transformation of the GfE Lagrangian. In fact we have
 \bea
 \mathcal{E}=\sum_{k=0}^4 z_k{\mathcal{G}}_ku_k-\mathcal{L},
 \eea
 which, using Eq.(\ref{LambdaG}) implies that
 \bea
 \mathcal{E}=\sum_{k=0}^4z_k[{\mathcal{G}_k}-{1}-\ln {\mathcal{G}_k}]=2\beta \Lambda_{\mathcal{G}}.
 \eea
  The internal energy density $\mathcal{E}$ of the GfE universes is small in  the low-energy, small-curvature limit $|u_k|=|1-\mathcal{G}_k^{-1}|\ll 1$ and vanishes at the first order approximation. However,  $\mathcal{E}$ diverges for $u_k=1-\mathcal{G}_k^{-1}\to 0$ or $u_k=1-\mathcal{G}_k^{-1}\to \infty$.
  
 {As we will see in the following, the GfE theory leads to a thermodynamic interpretation in which the geometric degrees of freedom are locally associated to notions of temperature and pressure. This thermal interpretation of the GfE theory might imply that the internal energy density $\mathcal{E}$ at each point in spacetime should be treated as the internal energy  of a thermal system interacting with the environment. Thus it should be associated to a contact manifold \cite{herczeg2018contact,bravetti2019contact,bravetti2017contact} rather than to the usual symplectic manifold associated to standard Hamiltonians. This interpretation will be rather crucial to establish the relation between the conjugated variables $\mathcal{G}_k$ and $\tau_k$ and ultimately providing key insights for the second quantization of the GfE theory. Further discussions of the implications of this observations are left for future works, while here we focus on the thermodynamics of the GfE theory.}

 {Building on the energy density $\mathcal{E}$ we} define  the total energy $\hat{\mathcal{E}}$ of spacetime is given by 
 \bea
 \bar{\mathcal{E}}=\frac{1}{\ell_P^4}\int \sqrt{-|g|}\mathcal{E }d^d{\bf r}.
 \eea
 The possible relation of the adopted  notion of internal energy density $\mathcal{E}$ with the Arnowitt-Deser-Misner (ADM) formalism \cite{arnowitt1960canonical,jha2023introduction} together with an in depth discussion of its associated  quantization are beyond the scope of this work that focuses on the thermodynamic aspects of the GfE theory and will be addressed in  subsequent publications.
\subsection{ Thermodynamics of GfE }
In order to formulate the thermodynamics of the GfE theory we  include in the treatment of the GfE action the notion of the {\em local volume} $\delta v$ defined as
 \bea
 \delta v=\frac{1}{\ell_P^4}\sqrt{-|g|},
 \eea
 which takes part in both the integral for the action $\mathcal{S}$ and the  $\bar{\mathcal{E}}$.
By including the volume contribution, and distinguishing between the spatial and temporal contributions of each order,  we define  {\em the local $k$-GQRE  } as

 \bea
\delta{s}_k=-\delta v \ln(1-u_k)=  \delta v\ln {\mathcal{G}}_k.
 \eea
 Thus we can express ${\mathcal{G}}_k$ in terms of the local volume and the local $k$-GQRE as
 \bea
 {\mathcal{G}}_k=e^{\delta s_k/ \delta v}.
 \eea
The {\em local $k$-energy} will be then derived directly from the internal energy density $\mathcal{E}$ of the GfE and defined as 
 \bea
 \delta \epsilon_k= \delta v\left[{\mathcal{G}}_k-1-\ln {\mathcal{G}}_k\right].
 \eea
 Having defined the local $k$-GQRE and the local $k$-energy we can determine their associated {\em $k$-temperature} $\theta_k$ and {\em $k$-pressure} $\pi_k$ using the thermodynamic relations
 \bea
 \frac{1}{\theta}_k=\left.\frac{\partial\delta{s}_k}{\partial  \delta\epsilon_k}\right|_{\delta v{=const}}\quad \frac{\pi_k}{\theta_k}=\left.\frac{\partial\delta{ s}_k}{\partial \delta v}\right|_{\delta\epsilon_k=const}.\label{tp}
 \eea
Note that  these notions of  temperatures and pressures are intrinsic to the geometric degrees of freedom of the GfE theory, coupling matter to geometry and depend non trivially on the G-field. Thus the $k$-temperatures are distinct from horizon temperatures associated with  quantum fields on curved-fixed background metrics.

From Eq.(\ref{tp}) it directly follows that the $k$-temperature can be expressed in terms of $\mathcal{G}_k$ or in terms of $u_k$ as
 \bea
 \theta_k={\mathcal{G}}_k-1=\frac{u_k}{1-u_k}.
 \eea
 For $|u_k|\ll 1$ we have that the $k$-temperature is well approximated by $u_k$  
 \bea
 \theta_k= u_k+O(u_k^2).
 \eea
  Note that in general the $k$-temperature $\theta_k$ can be both positive and negative depending on the relative sign of the curvature and the matter contributions. 
The $k$-pressure defined in Eq.(\ref{tp}) given by
 \bea
 \pi_k=\theta_k\frac{\delta s_k}{ \delta v}-\frac{ \delta\epsilon_k}{ \delta v}.
 \eea
 Therefore we obtain the first law of thermodynamics associated with the  GfE theory:
 \bea
 \delta\epsilon_k=\theta_k \delta s_k-\pi_k \delta v.
 \eea

 The thermodynamics of GfE opens new scenarios in gravity as it  implies that in the GfE universes are thermal as they are  locally associated with temperature and  pressure, which should be taken into account for a second quantization of this theory.  In particular  a quantization of the GfE theory will be needed to clarify if these temperatures are associated with  a radiation of particles  (gravitons).\\
These results are generally applicable to any solution of the GfE theory corresponding to homogeneous and isotropic spaces.
In order to provide a concrete example of how the thermodynamics of GfE might enrich our understanding of gravity, in the following we will  explore the value of the considered thermodynamic quantities drawing from insights provided by the Friedmann universes. It is to be emphasized that the Friedmann universes are solutions of the Einstein and Friedmann equations and thus will only be approximate solutions of the GfE equations of motion.
However  exploring  in the context of Friedmann universes the consequences deriving from the thermodynamics of GfE will reveal a new physical understanding of the thermodynamics of cosmologies.\\
\section{The Friedmann universes within the GfE theory}
In order  to address fundamental questions on the relation between the entropy and energy  of the universe   within the GfE theory, we adopt a simple but powerful approach that is common in theoretical physics when finding exact solutions of a statistical mechanics problem is challenging. In particular we will consider the Friedmann universes that are approximate solutions of the GfE action and insert the corresponding scalings of the physical quantities in the full GfE action. This will allow us  to extract the behaviour of the thermodynamic quantities of interest. This procedure is therefore similar to the treatment of critical phenomena \cite{negele2018quantum} such as the Ising model below the upper critical dimension leading to the Ginsburg criterion. In this latter context one desires to investigate the role of correlations in the full theory. To this end one inserts the mean-field solution (where these correlations are neglected) for estimating the correlations and explores to which extent the result is compatible with the assumption that correlations are negligible.
Similarly, here we will consider the Friedmann universes where one assumes that the G-field is given by the identity, i.e. $\tilde{\bm{\mathcal{G}}}=\tilde{\bf I}$  and we will use this solution to investigate the scalings of the thermodynamic quantities  of the full GfE theory. These include  the local entropy and the local energy that is non-trivial only if the G-field actually deviates from the identity. This approach  will allow us to determine under which conditions  the Friedmann universes remain good  approximations of the GfE universes.

In order to proceed along this direction, let us discuss in detail how Friedmann universes can be derived from the GfE theory.

Since we desire to consider the scenario in which the GfE equations at low coupling describe the Friedmann universe, we need to choose $\tilde{\bf M}$ in  a way  consistent with the description of  perfect fluids. In particular we impose that from the matter field Lagrangian $\mathcal{L}_M$  we   obtain the stress-energy tensor $T_{\mu\nu}$ in the corresponding Einstein equations, with
\bea
T_{\mu\nu}=(\rho+p)U_{\mu}U_{\nu}+pg_{\mu\nu},
\eea
where $\rho$ is the energy density and $p$ is the pressure of the perfect fluid.
This is achieved by considering 
\bea
{M}_{(1)}&=&\frac{1}{2}\left[\left(U_{\mu}U_{\nu}+\frac{1}{d}g_{\mu\nu}\right)\rho+\left(U_{\mu}U_{\nu}-\frac{1}{d}g_{\mu\nu}\right)p\right],\nonumber 
\eea
with the on shell constraint that $U_{\mu}U^{\mu}=-1$. Using the equation of state $p=w\rho$ we obtain
\bea
{M}_{(1)}&=&\frac{1}{2}\left(U_{\mu}U_{\nu}(1+w)+\frac{1}{d}g_{\mu\nu}(1-w)\right)\rho.
\eea
Using $U_{\mu}=(1,0,0,0)$ we thus get
\bea
{[{M}_{(1)}]}_{0}^{\ 0}&=&\frac{1}{2}\left(-(1+w)+\frac{1}{d}(1-w)\right)\rho,\nonumber\\
{[{M}_{(1)}]}_{i}^{\ j}&=&\delta_i^j\frac{1}{2d}(1-w)\rho,
\label{M}
\eea
while $M_{(0)}=0$ and $M_{(2)}=0$.
The value of $\rho$ is determined by the Friedmann equation
\bea
H^2=\frac{8\pi G}{3}\rho-\frac{\kappa}{a^2}
\eea
where $\kappa$ is the spatial curvature and $H$ is the Hubble parameter $H={\dot{a}}/{a}$.
The Friedmann solution for perfect fluids with $w\neq -1$ leads to the scaling
\bea
a\propto t^{2/n},\quad \rho\propto a^{-n},\quad \delta v=a^3 \propto t^{6/n},
\eea
where $n=3(1+w)$ where $w$ indicates the equation-of-state parameter depending on the nature of the fluid and determining the scaling of the pressure $p$ of the fluid and with the energy density $\rho$, i.e. $p=w\rho$. In particular we have
$n=4,w=1/3$ (radiation dominated); $n=3,w=0$ (matter dominated); $n=2,w=-1/3$ (curvature dominated).
In all these cases we obtain 
\bea
H=\left(\frac{2}{n}\right) t^{-1}.
\label{HF}
\eea
For the vacuum dominated fluid 
 $n=0,w=-1$ we have instead $H$ is constant, and
 \bea
 a\propto e^{Ht},\quad \rho\propto H^2,\quad \delta v=a^3\propto e^{3Ht}.
 \eea  
 Inserting these solutions in the Friedmann equation allows us to calculate $\rho$ as a function of $H$ for radiation, matter, curvature, and vacuum dominated universes.\\
 \begin{table*}
\begin{tabular}{||c|c cccc|c|c||}
\hline
\hline
FRW Cosmology  \qquad & $\omega_0/\beta $  \qquad & $\omega_1/\beta$ \qquad & $\omega_2/\beta $ \qquad & $\omega_3/\beta $\qquad & $\omega_4/\beta$ \qquad &$\bar{\omega}_{[1]}/\beta$ \qquad &$\bar{\omega}_{[2]}/\beta^2$\\
\hline
\hline
$\begin{array}{c}\mbox{General expression}\\
 w\neq-1/3,\kappa=0\end{array}$	 	& $3(1-3w)$ & $3(7+9w)/4$ & $3(w-1)/4$ & $-(1+3w)$  &$2$ & $9(1-w)$ &$ 3(71+42w+207 w^2)/8$\\
 \hline
$w=1/3,\kappa=0$  & $0$ & $15/2$ &$-1/2$ & $-2$ & $2$ & $6$ & $81/2$\\
 $w=0,\kappa=0$  &$3$ &$21/4$ &$-3/4$ & $-1$ & $2$ & $9$ & $213/8$\\
$w=-1,\kappa=0$  & $12$ & $-3/2$ &  $-3/2$ &$2$ &$2$ &$18$ & $177/2$\\
\hline
\hline
$w=-1/3,\kappa=-1,\rho=0$  & $0$ & $0$ & $0$ & $0$ & $0$&  $0$ & $0$\\
 \hline
 \hline
\end{tabular}
\caption{Values of the constant $\omega_k$,$\bar{\omega}_{[1]}$ and $\bar{\omega}_{[2]}$ defined in Eq.(\ref{omegak}) calculated for the Friedmann universes with $u_k$ defined according to Eq.(\ref{tauk}).}
\label{table_omega}
\end{table*}

\section{The thermodynamics of GfE  universes in the low-energy,  small-curvature limit}
 We can now calculate the thermodynamic properties of the GfE  homogeneous and isotropic universes in the limit of low energy and small curvature. Our strategy will be to neglect the effect due to high energy and large curvature by inserting the Friedmann universe solution into the GfE thermodynamic quantities. 
 As it happens for the mentioned parallelism with the Ising model, where the Ginsburg criterion indicates when fluctuations are no longer negligible, our strategy provides bounds for the validity of the Friedmann universes approximation to the GfE cosmology away from the  early universe. 
 Indeed, when we insert the scalings associated with the Friedmann universes into the definition of energy and entropy of the GfE theory we need to consider the solution for sufficiently large times so that $\tilde{\bf G}\tilde{g}^{-1}$  remains positively defined. This condition will guarantee us that we are considering the low energy, small curvature limit in which the GfE theory is well approximated by general relativity and will allow us to treat the dynamics away from the singularities of the general relativity.
In this situation, using the solution $\rho=\rho(t,n)$ and $a=a(t,n)$ of the Friedmann equations and the perfect fluid relation  $n=3(1+w)$ together with Eqs.(\ref{ab}), (\ref{M}) and inserting them into the definition of $u_k$ given by Eq.(\ref{tauk}) we get 
\bea
u_k=\omega_k H^2,
\label{omegak}
\eea
where $\omega_k=\omega_k^{\prime}\ell_P^2$ with $\omega_k^{\prime}$ given by the dimensionless constant, are given in Table $\ref{table_omega}$. Note that the thermodynamic scaling of $u_k$ captured by Eq.(\ref{omegak}) does not depend strongly on the matter model which only affects the value of $\omega_k$, and  the results hold for generic perfect fluids.
Indeed,  if we impose that the GfE Lagrangian is well defined, by requiring  $\tilde{\bf G}\tilde{g}^{-1}$ to be  positively defined for the Friedmann universes, we obtain that we can consider only times
\bea
t\gg t_0=\sqrt{\omega_{max}}
\eea
for $w\neq -1$ and Hubble constants
\bea
H\ll H_0=\frac{1}{\sqrt{\omega_{max}}}
\eea
for the de Sitter space $w=-1$,
where $\omega_{max}$ is given by 
\bea
\omega_{max}=\mbox{max}_{\omega_k>0}\omega_k.
\eea
Away from these limits, our approach is justifiable and we can explore safely the scaling of the thermodynamic quantities for the full GfE universes.

In this limit,  the local $k$-GQRE $\delta s_k$ the local $k$-energy $\delta \epsilon_k$ per unit volume are given by 
\bea
\delta s_k/\delta v&=&-\ln (1-\omega_k H^2),\nonumber \\
\delta \epsilon_k/\delta v&=& \frac{\omega_k H^2}{1-\omega_k H^2}+\ln (1-\omega_k H^2),\label{skek}\eea
while the $k$-temperature  $\theta_k$ and the $k$-pressure $\pi_k$ are given by 
\bea
\theta_k&=&\frac{\omega_k H^2}{1-\omega_k H^2},\nonumber \\
\pi_k&=&-\frac{1}{1-\omega_k H^2}\left[\omega_k H^2+\ln (1-\omega_k H^2)\right].
\eea
In the low-energy, small-curvature limit, $\ell_P H\ll1$ we obtain the scalings
\bea
\delta s_k/\delta v\simeq\omega_k H^2,\quad
\delta \epsilon_k/\delta v\simeq \frac{1}{2}\omega_k^2 H^4,\nonumber \\
\theta_k\simeq {\omega_k H^2},\quad \pi_k\simeq \frac{1}{2}{\omega_k^2 H^4}.\label{scaling_k}
\eea
Therefore in the considered limit, the $k$-temperature and the $k$-pressure for GfE universes approximated by the Friedmann universes different from the de Sitter space decay as $t^{-2}$ and $t^{-4}$ respectively, in the limit $t\gg 1$ while they are constant for GfE universe approximated by  the de Sitter space. 

The total local GQRE $\delta s$ and the total local energy $\delta \epsilon$  are given by
\bea
\delta s&=&\sum_{k=0}^4z_k\delta s_k,\quad
\delta \epsilon=\sum_{k=0}^4z_k\delta \epsilon_k
\label{set0}
\eea
 which in the limit of low energy $\ell_P H\ll 1$ scale like
\bea
\delta s/\delta v\simeq \bar{\omega}_{[1]} H^{2},\quad
\delta \epsilon/\delta v\simeq \bar{\omega}_{[2]} H^{4}
\eea
where the constants $\bar{\omega}_{[1]}$ and $\bar{\omega}_{[2]}$ are given by
\bea
\bar{\omega}_{[1]}&=&\sum_{k=0}^4 z_k \omega_k,\quad
{\bar\omega}_{[2]}=\frac{1}{2}\sum_{k=0}^4 z_k \omega_k^2.
\eea
For $w\neq-1$ we have then in the large time limit $t\gg 1$ the local GQRE and the local entropy per unit volume scale like
\bea
\delta s/\delta v\propto \bar{\omega}_{[1]} t^{-2},\quad
\delta \epsilon/\delta v\simeq \bar{\omega}_{[2]} t^{-4}.
\eea
Note that $\omega_k$ can have both positive and negative sign and as it is apparent from Table $\ref{table_omega}$, however
 $\bar{\omega}_{[1]}$, and $\bar{\omega}_{[2]}$ are positive for all Friedmann universes other than the Milne universe ($w=-1/3,n=2,\rho=0$) for which $\bar{\omega}_{[1]}=\bar{\omega}_{[2]}=0$. This implies that in the large time limit,  both the GQRE per unit volume and the energy per unit volume are positive and approach zero asymptotically.

  \begin{figure*}[!htb!]
  \includegraphics[width=1.9\columnwidth]{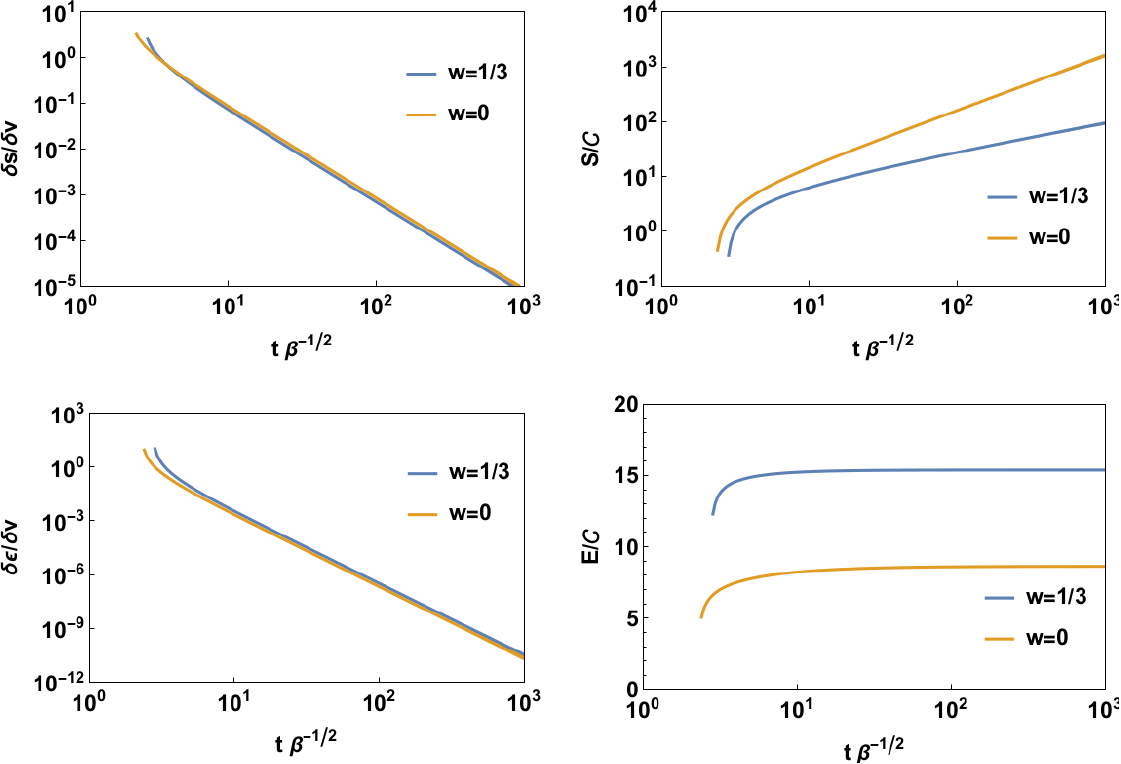}
  \caption{The thermodynamic properties of the GfE universes well  approximated by the radiation dominated ($w=1/3$) and the matter dominated Friedmann universes ($w=0$) as estimated from Eqs.(\ref{skek}), (\ref{set0}) and (\ref{SET}). The different panels display as a function of time $t$  the total local entropy per unit volume  $\delta s/\delta v$ (top left panel), the rescaled total entropy $S/\mathcal{C}$ (top right panel), the total local energy for unit volume $\delta \epsilon/\delta v$ (bottom left panel) and the rescaled total energy $E/\mathcal{C}$ (bottom right panel). Note that while the local entropy per unit volume $\delta s/\delta v$ decreases in time the total entropy $S$ increases in time. Similarly we observe that while the local energy per unit volume $\delta\epsilon/\delta v$ decreases in time, the total energy $E$ is not decreasing and saturates to a constant value.}
  \label{fig2}
 \end{figure*}

For GfE universes well approximated by Friedmann universes different from the de Sitter space, the total entropy $S$ and the total energy $E$   are given respectively by $\delta s$ and $\delta \epsilon$ integrated  over the spacetime region $ (t_0,t)\times \mathcal{V}$, where $\mathcal{V}$ indicates a spatial region of spacetime. Specifically, the total entropy $S$ and the total energy $E$ of GfE universes  are defined as
\bea
S=\mathcal{V}\int_{t_0}^t d\tau  \delta s,\quad
E=\mathcal{V}\int_{t_0}^t d\tau  \delta \epsilon.
\label{SH}  
\eea
Considering the expression in Eq.(\ref{skek})  we obtain the following expressions for $S$ and $E$, 
\bea
S&=&-\mathcal{C}\int_{t_0}^t \left(\frac{\tau}{t_0}\right)^{6/n}\sum_{k=0}^4z_k\ln (1-\omega_k \tau^{-2}) d\tau,\label{SET} \\
E&=& \mathcal{C}\int_{t_0}^t \left(\frac{\tau}{t_0}\right)^{6/n}\sum_{k=0}^4 z_k \left[\frac{\omega_k \tau^{-2}}{1-\omega_k \tau^{-2}}+\ln (1-\omega_k \tau^{-2})\right]d\tau,\nonumber \eea
where $\mathcal{C}=(a(t_0))^3\mathcal{V}.$
Using for  $\delta s_k$ and $\delta \epsilon_k$ the scaling relations in Eq.(\ref{scaling_k})  and using Eq.(\ref{HF}) for expressing the $H$ as a function of $t$, we obtain that in the large times $t\gg 1$ limit $S$ and $E$ scale like 
\bea
S\propto  \bar{\omega}_{[1]} t^{6/n-1},\quad E  \propto  \bar{\omega}_{[2]}t^{6/n-3},
\eea
 as long as $w\neq-1, n\neq 0$.  Specifically, noting that the Milne universe $n=2$ has $\omega_k=0$ for every value of $k$, we obtain, for the total entropy $S$ 
\bea
{S}/\bar{\omega}_{[1]}\propto\left\{\begin{array}{lll}t^{1/2}&\mbox{for} &n=4,\\
t&\mbox{for} & n=3,\\
0&\mbox{for} &n=2,
\end{array}\right.
\eea
while for the total energy $E$, in leading order of $t$ scales like 
\bea
E/\bar{\omega}_{[2]}\propto\left\{\begin{array}{lll}\mbox{const}&\mbox{for} &n\in \{3,4\}.\\
0 &\mbox{for} &n=2.
\end{array}\right.
\eea
Thus, for the physically relevant cases of matter and radiation dominated universes  the total entropy $S$ increases in time, due to the effect of the increasing volume while the total energy is in first approximation constant.
Interestingly, similar results can be obtained with the due modifications, for the scaling of the total entropy and total energy in terms of the conformal time.
For the de Sitter space, both the total entropy and the total energy defined as in Eq.(\ref{SH}) increase in time exponentially, however for a single observer not all spacetime is observable so we might want to define the total entropy and the total energy as integrated over the causal diamond of de Sitter space $\mathcal{K}_{dS}$. According to this definition we have 
\bea
S=\frac{1}{\ell_P^4}\int_{\mathcal{K}_{dS}}\sqrt{-|g|}\mathcal{L} d^4{\bf r},\quad E=\frac{1}{\ell_P^4}\int_{\mathcal{K}_{dS}}\sqrt{-|g|}\mathcal{E}d^4{\bf r},\nonumber 
\eea
Therefore indicating with ${\mathcal{V}_{dS}}\simeq H^{-4}$ the finite causal diamond volume  we obtain
\bea
S&=& \frac{\mathcal{V}_{dS}}{\ell_P^4}\mathcal{L}\simeq\frac{\delta s}{\delta v}\frac{1}{\ell_P^4 H^4},\nonumber \\
E&=&\frac{\mathcal{V}_{dS}}{\ell_P^4}\mathcal{E}\simeq \frac{\delta \epsilon}{\delta v}\frac{1}{\ell_P^4 H^4}.
\eea
In the limit  $\ell_P H\ll 1$  we obtain
\bea
S& \simeq \frac{\bar{\omega}_{[1]}}{\ell_P^4H^2},\quad E&\simeq \frac{{\bar{\omega}}_{[2]}}{\ell_P^4}.
\eea 
Therefore the total entropy scales like $S=O(H^{-2})$ in agreement with the scaling of the Gibbons-Hawking entropy \cite{gibbons1977cosmological} for de Sitter space, and the total energy $E=\mathcal{O}(1)$. Note however that  the $k$-temperature $\theta_k$ associated with de Sitter space is given by Eq.(\ref{scaling_k}) and thus  obeys a different scaling from the Gibbons-Hawking and Bunch-Davies temperature \cite{bunch1978quantum,gibbons1977cosmological} of de Sitter space as we have $\theta_k\propto \omega_k H^2$. This different scaling reveals that  the $k$-temperature is inherently associated with the geometry spacetime rather than the quantum fields defined on it. This might imply that the $k$-temperature might induce  the radiation of gravitons rather than the emission of particles associated with the quantum fields defined on  the de Sitter background. \\
\section{ Conclusions}

 {In this work, we have uncovered the thermodynamics of Gravity from Entropy (GfE) encoding the information  in the interplay between spacetime geometry and matter fields. While our analysis is developed within the specific GfE framework, the results we obtain are of  relevance for GR as well,  {as far as GR is interpreted as  the low energy small curvature limit of the GfE theory.}}

 {Most of the existing approaches to gravitational thermodynamics rely crucially on horizon entropy and local horizon constructions. While highly successful, they do not directly address how local gravitational ordering, structure formation, and complexity can coexist with the second law of thermodynamics. In contrast, the GfE framework enables the study of the thermodynamic aspects of gravity and its associated cosmological evolution independently of the presence of horizons.}

 {Moreover, so far in  GR entropy is primarily attributed to either matter, radiation, or to  horizons, and geometric degrees of freedom. In contrast, the GfE framework encodes the information  the interplay between the geometric  degrees and matter fields.}

 {By focusing on homogeneous and isotropic spacetimes described by the FRW metric, in this work we have identified the GfE energy density with the  {emergent effective dark energy term $\Lambda_{\mathcal{G}}$.} We have shown that GfE universes are thermal, in the sense that their metric degrees of freedom are associated with $k$-temperatures and $k$-pressures, depending on their order and on their spacelike or timelike nature. The resulting thermodynamics reveals that, within the GfE theory, locally, spacetime can be interpreted as a thermodynamic system describing the interaction between the geometric degrees of freedom and the matter fields. Each of these thermodynamic systems  is interacting with the environment and satisfies the  first law of GfE thermodynamics.} 

 { {Although the GfE theory is still in its infancy and needs to be validated and tested against cosmological and astrophysical observations, the  GfE thermodynamics presented here  provides  new insight into the possible  relationship between gravity and the second law of thermodynamics.} In FRW spacetimes,  {when the GfE equation of motion are well approximated by GR,} we find that the total spacetime entropy is non-decreasing in time while the local entropy density can decrease, showing that cosmological dynamics can accommodate  local ordering while remaining fully consistent with the second law.  {Therefore the present manuscript effectively reinterprets GR, and specifically the Friedmann universes in thermodynamic terms, provided GR is considered as the low energy small curvature limit of the GfE theory.}
More broadly, by embedding gravitational dynamics within a thermodynamic and information-theoretic framework, GfE opens new avenues for investigating  {the long-standing problem of reconciling } the foundations of cosmological irreversibility, the emergence of complex structures, and ultimately life,  with fundamental gravitational dynamics.}

 {Our results also open new perspectives for quantum gravity. Since the GfE thermodynamics exhibit an intrinsic thermal character, a possible quantization of the theory may naturally require a formulation based on contact geometry, associated with thermodynamic states, rather than on symplectic geometry alone, which is traditionally tied to conservative Hamiltonian dynamics.  {Thus this result provides a change of perspective which  brings new insights into a possible route for quantization of the GfE theory}.}

 {Finally, we note that the GfE action yields a description of local degrees of freedom, encompassing both geometry and matter fields, as thermal and thus, open thermodynamic systems interacting with the environment. This notable feature of the GfE thermodynamics opens broad perspectives in relation to the interpretation of  the GfE action as  entanglement entropy and in the establishment of the full gravitational consequences of this interpretation.}
 
{In summary, the GfE thermodynamics derived here  opens new perspectives for  theories of classical and quantum gravity, statistical mechanics, the theory of entanglement, and cosmology.}

\bibliographystyle{unsrt}
\bibliography{references}
\end{document}